\documentclass[10pt]{article} 
\usepackage{a4wide}

\makeatletter

\usepackage{amstext}
\usepackage{amssymb}
\usepackage{stmaryrd}
\DeclareFontFamily{OT1}{cmtex}{}
\DeclareFontShape{OT1}{cmtex}{m}{n}
  {<5><6><7><8>cmtex8
   <9>cmtex9
   <10><10.95><12><14.4><17.28><20.74><24.88>cmtex10}{}
\DeclareFontShape{OT1}{cmtex}{m}{it}
  {<-> ssub * cmtt/m/it}{}

\DeclareFontShape{OT1}{cmtt}{bx}{n}
  {<5><6><7><8>cmtt8
   <9>cmbtt9
   <10><10.95><12><14.4><17.28><20.74><24.88>cmbtt10}{}
\DeclareFontShape{OT1}{cmtex}{bx}{n}
  {<-> ssub * cmtt/bx/n}{}

\newcommand{\Varid}[1]{\mathit{#1}}
\newcommand{\anonymous}{\kern0.06em \vbox{\hrule\@width.5em}}

\usepackage{polytable}

\@ifundefined{mathindent}%
  {\newdimen\mathindent\mathindent\leftmargini}%
  {}%

\def\resethooks{%
  \global\let\SaveRestoreHook\empty
  \global\let\ColumnHook\empty}
\newcommand*{\savecolumns}[1][default]%
  {\g@addto@macro\SaveRestoreHook{\savecolumns[#1]}}
\newcommand*{\restorecolumns}[1][default]%
  {\g@addto@macro\SaveRestoreHook{\restorecolumns[#1]}}
\newcommand*{\aligncolumn}[2]%
  {\g@addto@macro\ColumnHook{\column{#1}{#2}}}

\resethooks

\newcommand{\onelinecommentchars}{\quad-{}- }
\newcommand{\commentbeginchars}{\enskip\{-}
\newcommand{\commentendchars}{-\}\enskip}

\newcommand{\visiblecomments}{%
  \let\onelinecomment=\onelinecommentchars
  \let\commentbegin=\commentbeginchars
  \let\commentend=\commentendchars}

\newcommand{\invisiblecomments}{%
  \let\onelinecomment=\empty
  \let\commentbegin=\empty
  \let\commentend=\empty}

\visiblecomments

\newlength{\blanklineskip}
\newcommand{\hsindent}[1]{\quad}

\makeatother

\input{Qcircuit}
\newdimen\proofrulebreadth \proofrulebreadth=.05em
\newdimen\proofdotseparation \proofdotseparation=1.25ex
\newdimen\proofrulebaseline \proofrulebaseline=2ex
\newcount\proofdotnumber \proofdotnumber=3
\let\then\relax
\def\hfi{\hskip0pt plus.0001fil}
\mathchardef\squigto="3A3B
\newif\ifinsideprooftree\insideprooftreefalse
\newif\ifonleftofproofrule\onleftofproofrulefalse
\newif\ifproofdots\proofdotsfalse
\newif\ifdoubleproof\doubleprooffalse
\let\wereinproofbit\relax
\newdimen\shortenproofleft
\newdimen\shortenproofright
\newdimen\proofbelowshift
\newbox\proofabove
\newbox\proofbelow
\newbox\proofrulename
\def\shiftproofbelow{\let\next\relax\afterassignment\setshiftproofbelow\dimen0 }
\def\shiftproofbelowneg{\def\next{\multiply\dimen0 by-1 }%
\afterassignment\setshiftproofbelow\dimen0 }
\def\setshiftproofbelow{\next\proofbelowshift=\dimen0 }
\def\setproofrulebreadth{\proofrulebreadth}

\def\prooftree{
\ifnum  \lastpenalty=1
\then   \unpenalty
\else   \onleftofproofrulefalse
\fi
\ifonleftofproofrule
\else   \ifinsideprooftree
        \then   \hskip.5em plus1fil
        \fi
\fi
\bgroup
\setbox\proofbelow=\hbox{}\setbox\proofrulename=\hbox{}%
\let\justifies\proofover\let\leadsto\proofoverdots\let\Justifies\proofoverdbl
\let\using\proofusing\let\[\prooftree
\ifinsideprooftree\let\]\endprooftree\fi
\proofdotsfalse\doubleprooffalse
\let\thickness\setproofrulebreadth
\let\shiftright\shiftproofbelow \let\shift\shiftproofbelow
\let\shiftleft\shiftproofbelowneg
\let\ifwasinsideprooftree\ifinsideprooftree
\insideprooftreetrue
\setbox\proofabove=\hbox\bgroup$\displaystyle 
\let\wereinproofbit\prooftree
\shortenproofleft=0pt \shortenproofright=0pt \proofbelowshift=0pt
\onleftofproofruletrue\penalty1
}

\def\eproofbit{
\ifx    \wereinproofbit\prooftree
\then   \ifcase \lastpenalty
        \then   \shortenproofright=0pt  
        \or     \unpenalty\hfil         
        \or     \unpenalty\unskip       
        \else   \shortenproofright=0pt  
        \fi
\fi
\global\dimen0=\shortenproofleft
\global\dimen1=\shortenproofright
\global\dimen2=\proofrulebreadth
\global\dimen3=\proofbelowshift
\global\dimen4=\proofdotseparation
\global\count255=\proofdotnumber
$\egroup  
\shortenproofleft=\dimen0
\shortenproofright=\dimen1
\proofrulebreadth=\dimen2
\proofbelowshift=\dimen3
\proofdotseparation=\dimen4
\proofdotnumber=\count255
}

\def\proofover{
\eproofbit 
\setbox\proofbelow=\hbox\bgroup 
\let\wereinproofbit\proofover
$\displaystyle
}%
\def\proofoverdbl{
\eproofbit 
\doubleprooftrue
\setbox\proofbelow=\hbox\bgroup 
\let\wereinproofbit\proofoverdbl
$\displaystyle
}%
\def\proofoverdots{
\eproofbit 
\proofdotstrue
\setbox\proofbelow=\hbox\bgroup 
\let\wereinproofbit\proofoverdots
$\displaystyle
}%
\def\proofusing{
\eproofbit 
\setbox\proofrulename=\hbox\bgroup 
\let\wereinproofbit\proofusing
\kern0.3em$
}

\def\endprooftree{
\eproofbit 
  \dimen5 =0pt
\dimen0=\wd\proofabove \advance\dimen0-\shortenproofleft
\advance\dimen0-\shortenproofright
\dimen1=.5\dimen0 \advance\dimen1-.5\wd\proofbelow
\dimen4=\dimen1
\advance\dimen1\proofbelowshift \advance\dimen4-\proofbelowshift
\ifdim  \dimen1<0pt
\then   \advance\shortenproofleft\dimen1
        \advance\dimen0-\dimen1
        \dimen1=0pt
        \ifdim  \shortenproofleft<0pt
        \then   \setbox\proofabove=\hbox{%
                        \kern-\shortenproofleft\unhbox\proofabove}%
                \shortenproofleft=0pt
        \fi
\fi
\ifdim  \dimen4<0pt
\then   \advance\shortenproofright\dimen4
        \advance\dimen0-\dimen4
        \dimen4=0pt
\fi
\ifdim  \shortenproofright<\wd\proofrulename
\then   \shortenproofright=\wd\proofrulename
\fi
\dimen2=\shortenproofleft \advance\dimen2 by\dimen1
\dimen3=\shortenproofright\advance\dimen3 by\dimen4
\ifproofdots
\then
        \dimen6=\shortenproofleft \advance\dimen6 .5\dimen0
        \setbox1=\vbox to\proofdotseparation{\vss\hbox{$\cdot$}\vss}%
        \setbox0=\hbox{%
                \advance\dimen6-.5\wd1
                \kern\dimen6
                $\vcenter to\proofdotnumber\proofdotseparation
                        {\leaders\box1\vfill}$%
                \unhbox\proofrulename}%
\else   \dimen6=\fontdimen22\the\textfont2 
        \dimen7=\dimen6
        \advance\dimen6by.5\proofrulebreadth
        \advance\dimen7by-.5\proofrulebreadth
        \setbox0=\hbox{%
                \kern\shortenproofleft
                \ifdoubleproof
                \then   \hbox to\dimen0{%
                        $\mathsurround0pt\mathord=\mkern-6mu%
                        \cleaders\hbox{$\mkern-2mu=\mkern-2mu$}\hfill
                        \mkern-6mu\mathord=$}%
                \else   \vrule height\dimen6 depth-\dimen7 width\dimen0
                \fi
                \unhbox\proofrulename}%
        \ht0=\dimen6 \dp0=-\dimen7
\fi
\let\doll\relax
\ifwasinsideprooftree
\then   \let\VBOX\vbox
\else   \ifmmode\else$\let\doll=$\fi
        \let\VBOX\vcenter
\fi
\VBOX   {\baselineskip\proofrulebaseline \lineskip.2ex
        \expandafter\lineskiplimit\ifproofdots0ex\else-0.6ex\fi
        \hbox   spread\dimen5   {\hfi\unhbox\proofabove\hfi}%
        \hbox{\box0}%
        \hbox   {\kern\dimen2 \box\proofbelow}}\doll%
\global\dimen2=\dimen2
\global\dimen3=\dimen3
\egroup 
\ifonleftofproofrule
\then   \shortenproofleft=\dimen2
\fi
\shortenproofright=\dimen3
\onleftofproofrulefalse
\ifinsideprooftree
\then   \hskip.5em plus 1fil \penalty2
\fi
}

\newcommand{\FinSet}{\mathbf{FinSet}}
\newcommand{\Id}{\mathrm{I}}
\newcommand{\FCC}{\mathbf{FCC}}
\newcommand{\FCCo}{\mathbf{FCC}^\circ}
\newcommand{\FCCm}[2]{\FCC\,#1\,#2}
\newcommand{\forget}[1]{\textrm{U}_{#1}}

\newcommand{\forgetC}{\forget{\FCC}}
\newcommand{\FQC}{\mathbf{FQC}}
\newcommand{\FQCo}{\mathbf{FQC}^\circ}
\newcommand{\FQCm}[2]{\FQC\,#1\,#2}
\newcommand{\FQCom}[2]{\FQCo\,#1\,#2}
\newcommand{\forgetQ}{\forget{\FQC}}
\newcommand{\exteq}{=_\textrm{ext}}
\newcommand{\CVec}[1]{\Complex^{#1}}

\newcommand{\LinArr}[2]{#1\multimap #2}
\newcommand{\UnitaryArr}[2]{#1\multimap_{\textrm{unitary}} #2}

\newcommand{\Dens}[1]{\mathrm{Dens}\,{#1}}
\newcommand{\Super}{\mathbf{Super}}
\newcommand{\SuperArr}[2]{#1\multimap_{\textrm{super}} #2}
\newcommand{\Complex}{\mathbb{C}}

\newcommand{\adjoin}[1]{\widehat{#1}}
\newcommand{\lift}[1]{\widetilde{#1}}

\newcommand{\adj}[1]{{#1}^\dagger}

\newcommand{\tr}{\mathrm{tr}}

\newcommand{\QQ}[1]{{\cal Q}_{#1}}
\newcommand{\Qbit}{{\QQ{2}}}

\newcommand{\ot}{\otimes}
\newcommand{\emptyC}{\bullet}
\newcommand{\G}{\Gamma}
\newcommand{\D}{\Delta}
\newcommand{\dom}[1]{\textrm{dom}\,{#1}}
\newcommand{\den}[1]{\llbracket #1 \rrbracket}

\newcommand{\true}{\mathrm{true}}
\newcommand{\false}{\mathrm{false}}
\newcommand{\CO}{\textrm{C}}
\newcommand{\WK}{\textrm{W}}
\newcommand{\PAD}{\textrm{P}}
\newcommand{\choice}[2]{[#1|#2]}

\newcommand{\ceil}[1]{\lceil{#1}\rceil}
\newcommand{\vdasho}{\vdash^\circ}

\newcommand{\ru}[2]{\vspace{1ex}
\begin{prooftree}
#1 \justifies #2
\end{prooftree}\vspace{1ex}}

\newcommand{\Ru}[3]{\vspace{1ex}
\begin{prooftree}
#1 \justifies #2 \using{\rm{#3}}
\end{prooftree}\vspace{1ex}}
\newcommand{\Ax}[2]{
\Ru{}{#1}{#2} }

\newcommand{\qtrue}{\mathtt{qtrue}}
\newcommand{\qfalse}{\mathtt{qfalse}}
\newcommand{\rin}{\ \mathtt{in}\ }

\newcommand{\rifo}{\ \mathtt{if}^{\circ}\ \ }
\newcommand{\rthen}{\ \ \mathtt{then}\ \ }
\newcommand{\relse}{\ \ \mathtt{else}\ \ }
\newcommand{\rlet}{\mathtt{let}\ }
\newcommand{\rinl}{\mathtt{inl}\ }
\newcommand{\rinr}{\mathtt{inr}\ }
\newcommand{\rcase}{\mathtt{case}\ }
\newcommand{\rcaseo}{\mathtt{case}^\circ\ }
\newcommand{\rof}{\,\,\mathtt{of}\ \ }
\newcommand{\abs}[1]{\vert #1 \vert}
\newcommand{\tin}{:}

\newcommand{\lab}[1]{\pushT{#1}} 
\newcommand{\Cgg}{\multigate{1}{\phi_{C}}}
\newcommand{\gCgg}{\ghost{\phi_{C}}}
\newcommand{\sut}{\sigma\sqcup\tau}

\newcommand{\Tyid}[1]{\mathbf{#1}}
\newcommand{\Cnid}[1]{\mathrm{#1}}

\begin{document}

\title{A functional quantum programming language}

\author{Thorsten Altenkirch and Jonathan Grattage\\
  School of Computer Science and IT, Nottingham University\\
  email: \texttt{\{txa,jjg\}@cs.nott.ac.uk}}

\date{February 2005}

\maketitle

\begin{abstract}
  We introduce the language QML, a functional language for quantum
  computations on finite types. Its design is guided by its
  categorical semantics: QML programs are interpreted by morphisms in
  the category $\FQC$ of finite quantum computations, which provides a
  constructive semantics of irreversible quantum computations
  realisable as quantum gates.  QML integrates reversible and
  irreversible quantum computations in one language, using first order
  strict linear logic to make weakenings explicit. Strict programs are
  free from decoherence and hence preserve superpositions and entanglement -- which is
  essential for quantum parallelism.
\end{abstract}

\section{Introduction}
\label{sec:intro}


The discovery of efficient quantum algorithms by Shor \cite{Shor94}
and Grover \cite{Grov97} has triggered much interest in the field of
quantum programming. However, it is still a very hard task to find
new quantum algorithms. One of the reasons for this situation might
be that quantum programs are very low level: they are usually
represented as quantum circuits, or in some combinator language
which gives rise to circuits. Here we attempt to remedy this
situation by introducing the quantum programming language QML, which
is based on high-level constructs known from conventional functional
programming. Though functional (programs are expressions), our
language is first order and finitary; all datatypes are finite. We
will discuss possible extensions in the conclusions, but we believe
that the approach presented here represents a significant progress
towards the goal of a natural quantum programming language.

We present a semantics of our language by interpreting terms as morphisms
in the category of finite quantum computations $\FQC$, which we introduce here.
The $\FQC$ semantics gives rise to a denotational semantics in terms
of superoperators, the accepted domain of irreversible quantum computation, and
at the same time to a compiler into quantum circuits, an accepted operational
semantics for quantum programs.

As an illustration, one of the basic quantum circuits is the
Hadamard gate, which is usually defined by presenting its matrix:
\[ \mathrm{had} =  \frac{1}{\sqrt{2}} \begin{pmatrix}
  1 & 1 \\    1 & -1 \\
  \end{pmatrix}\]
But what does this mean in programming terms? In QML this operation is
implemented by the following program

\begingroup\par\noindent\advance\leftskip\mathindent\(
\begin{pboxed}\SaveRestoreHook
\column{B}{@{}l@{}}
\column{3}{@{}l@{}}
\column{12}{@{}l@{}}
\column{18}{@{}l@{}}
\column{E}{@{}l@{}}
\fromto{3}{E}{{}\Varid{had}\mathbin{:}\Tyid{Q_2}\multimap\Tyid{Q_2}{}}
\nextline
\fromto{3}{12}{{}\Varid{had}\;\Varid{x}\mathrel{=}{}}
\fromto{12}{E}{{}\mathbf{if}^\circ\;\Varid{x}{}}
\nextline
\fromto{12}{18}{{}\mathbf{then}\;{}}
\fromto{18}{E}{{}\{\mskip1.5mu \Cnid{qfalse}\mid (\mathbin{-}\mathrm{1})\;\Cnid{qtrue}\mskip1.5mu\}{}}
\nextline
\fromto{12}{18}{{}\mathbf{else}\;{}}
\fromto{18}{E}{{}\{\mskip1.5mu \Cnid{qfalse}\mid \Cnid{qtrue}\mskip1.5mu\}{}}
\ColumnHook
\end{pboxed}
\)\par\noindent\endgroup\resethooks

We can read \ensuremath{\Varid{had}} as an operation which, depending on its input qubit
$x$, returns one of two superpositions of a qubit. We can also easily
calculate that applying \ensuremath{\Varid{had}} twice gets us back where we started by
cancelling out amplitudes.

An important feature of quantum programming is the possibility to
create superpositions which have non-local effects. A simple
application of this idea is the algorithm in figure
\ref{fig:deutsch} to determine whether two classical bits, represented as
qubits, are equal, which is based on Deutsch's algorithm (see
\cite{NC00}, pp.32). It exploits quantum parallelism by querying
both inputs at the same time; this corresponds to the fact that the
expressions \ensuremath{\mathbf{if}^\circ\;\Varid{a}} and \ensuremath{\mathbf{if}^\circ\;\Varid{b}} in our program are not nested. The
famous algorithms by Shor and Grover rely on a more subtle
exploitation of this effect.

\begin{figure*}[htbp]
\begingroup\par\noindent\advance\leftskip\mathindent\(
\begin{pboxed}\SaveRestoreHook
\column{B}{@{}l@{}}
\column{12}{@{}l@{}}
\column{25}{@{}l@{}}
\column{40}{@{}l@{}}
\column{E}{@{}l@{}}
\fromto{B}{E}{{}\Varid{eq}\mathbin{:}\Tyid{Q_2}\multimap\Tyid{Q_2}\multimap\Tyid{Q_2}{}}
\nextline[\blanklineskip]
\fromto{B}{12}{{}\Varid{eq}\;\Varid{a}\;\Varid{b}\mathrel{=}{}}
\fromto{12}{25}{{}\mathbf{let}\;(\Varid{x},\Varid{y})\mathrel{=}{}}
\fromto{25}{E}{{}\mathbf{if}^\circ\{\mskip1.5mu \Cnid{qfalse}\mid \Cnid{qtrue}\mskip1.5mu\}{}}
\nextline
\fromto{25}{40}{{}\mathbf{then}\;(\Cnid{qtrue},{}}
\fromto{40}{E}{{}\mathbf{if}^\circ\;\Varid{a}{}}
\nextline
\fromto{40}{E}{{}\mathbf{then}\;(\{\mskip1.5mu \Cnid{qfalse}\mid (\mathbin{-}\mathrm{1})\;\Cnid{qtrue}\mskip1.5mu\},(\Cnid{qtrue},\Varid{b})){}}
\nextline
\fromto{40}{E}{{}\mathbf{else}\;(\{\mskip1.5mu (\mathbin{-}\mathrm{1})\;\Cnid{qfalse}\mid \Cnid{qtrue}\mskip1.5mu\},(\Cnid{qfalse},\Varid{b}))){}}
\nextline
\fromto{25}{40}{{}\mathbf{else}\;(\Cnid{qfalse},{}}
\fromto{40}{E}{{}\mathbf{if}^\circ\;\Varid{b}{}}
\nextline
\fromto{40}{E}{{}\mathbf{then}\;(\{\mskip1.5mu (\mathbin{-}\mathrm{1})\;\Cnid{qfalse}\mid \Cnid{qtrue}\mskip1.5mu\},(\Varid{a},\Cnid{qtrue})){}}
\nextline
\fromto{40}{E}{{}\mathbf{else}\;(\{\mskip1.5mu \Cnid{qfalse}\mid (\mathbin{-}\mathrm{1})\;\Cnid{qtrue}\mskip1.5mu\},(\Varid{a},\Cnid{qfalse}))){}}
\nextline
\fromto{12}{E}{{}\mathbf{in}\;\Varid{had}\;\Varid{x}{}}
\ColumnHook
\end{pboxed}
\)\par\noindent\endgroup\resethooks
\centering \caption{A variant of Deutsch's algorithm}
\label{fig:deutsch}
\end{figure*}

The reader may have noticed that we do not insist on quantum programs
being reversible. We will discuss this further in section
\ref{sec:fcqc}, by comparing classical and quantum computation.  It
turns out that in both cases irreversible computations can be reduced
to reversible ones in a similar fashion. However, reversibility plays
a more central role in quantum computation due to the fact that
forgetting information leads to decoherence, which destroys
entanglement, and hence negatively affects quantum parallelism. Thus
one of the central features of our language is \emph{control of
  decoherence}, which is achieved by keeping track of weakening
through the use of strict linear logic (or just strict logic) and by
offering different if-then-else (or, generally, case) operators, one
that measures the qubit, \ensuremath{\mathbf{if}}, and a second, \ensuremath{\mathbf{if}^\circ}, that doesn't -- but
which can only be used in certain situations. We hasten to add that
this \emph{intrinsic decoherence} is not related to the decoherence
which is caused by thermal noise in a hypothetical quantum computer.
As one of the referees has pointed out, \emph{control of decoherence} is
in spirit similar to Reynold's \emph{control of interference}
\cite{reynolds}.

\section{Related work}
\label{sec:related}

There are a number of papers on simulating or integrating quantum
programming within conventional functional programming, e.g.  Mu and
Bird's proposal on modelling quantum programming in a functional
language \cite{BirdMu02}, Karczmarczuk's use of functional programming to
model quantum systems \cite{Karc03} and Sabry's proposal to structure
embedded quantum programs using \emph{virtual values} \cite{Sabry03}.
Yet another approach was suggested by Sanders and Zuliani
\cite{Zuliani00}, which extends the probabilistic guarded command
language \cite{pgcl} by quantum registers and operations on quantum
registers.

Peter Selinger's influential paper \cite{selinger:qpl} introduces a
single-assignment (essentially functional) quantum programming
language, which is based on the separation of \emph{classical control}
and \emph{quantum data}. This language combines high-level classical
structures with operations on quantum data, and has a clear
mathematical semantics in the form of superoperators. Quantum data can
be manipulated by using unitary operators or by measurement, which can
affect the classical control flow.  Recently, Selinger and Valiron
\cite{selingerValiron:lambda} have presented a functional language
based on the \emph{classical control} and \emph{quantum data} paradigm.

Selinger and Valiron's approach is in some sense complementary
to ours: they use an affine type system (no contraction), while we
use a strict system (no weakening). The lack of contraction is
justified by the no-cloning property of quantum
states. However, this does not apply to our approach as we model
contraction by sharing, not by copying - this is also used in \cite{arrighiDowek}.
Indeed, classical programming languages do not implement
contraction by copying data, but by sharing via pointers.

Andre van Tonder has proposed a quantum $\lambda$-calculus
incorporating higher order \cite{tonder1,tonder2} programs, however he
is not considering measurements as part of his language. In \cite{tonder2}
he suggests a semantics for a strict higher order quantum language
based on vector bundles. At the current time it is not clear, to us,
whether the details of this construction work out.

Abramsky and Coecke \cite{AbramCoecke04} have investigated a
categorical semantics for quantum protocols using the compact
closed structure of the category of finite dimensional Hilbert spaces.
They suggest that their semantics may be relevant for type systems
for quantum programming language. It remains to be seen how
this relates to our work, since our approach does not exploit
compact closure.

All the previous approaches adopt a basically combinatory approach to
quantum data: operations on quantum data are given by combinators
implementing unitary operators. We believe that our work is novel in
that we are proposing high-level quantum control structures, i.e. we
are aiming at quantum control and quantum data.

\section{Finite classical and quantum computation}
\label{sec:fcqc}

It is frequently emphasised that quantum computation relies on
reversibility because quantum physics models reversible processes.
This is true, but the same holds for classical computation ---
whether we base our notion of computation on Newtonian physics or
Maxwellian electrodynamics, the underlying physical processes are
reversible for a closed system. Hence we should explain irreversible classical
computation based on a reversible mechanism. Here, we will develop a
picture which applies to classical and quantum computation.  This
makes it easy to identify the essential differences and also guides
the design of QML which realises structures common to both
computational paradigms by syntactic constructs established in
classical functional programming.

We introduce the category $\FQC$ of finite quantum computations and,
for purposes of comparison, the category $\FCC$ of finite classical
computations%
\footnote{$\FCC$ may be viewed as a categorical account of a finite
version of Bennet's results \cite{benn:lrc73w}.}%
. We will interpret QML programs by $\FQC$ morphisms. It
is straightforward to identify a classical sublanguage of QML
which can be interpreted in $\FCC$; however we will not carry this
out in detail.

Objects of both categories are finite sets, for which we use the
letters $A,B,C$. While classical computations are carried out on the
elements of those sets, quantum computations take place in finite
dimensional Hilbert spaces; we write $\CVec{A}$ for the space
generated by $A$, whose elements are functions \footnote{$\CVec{A}$
gives rise to a \emph{Kleisli structure}, \cite{alti:csl99}, here
\textrm{bind} is realised by matrix multiplication. Its Kleisli
category is the category of finite dimensional vector spaces.}.
  A reversible finite computation, that is a closed
computational system, is modelled by a reversible operation $\phi$,
which is a bijection of finite sets in the classical case, and a
unitary operator on the Hilbert spaces in the quantum case. We write
$\UnitaryArr{A}{B}$ for the set of unitary operators from the space 
generated by $A$ to the space generated by $B$, which in the
finite-dimensional case correspond exactly to norm-preserving linear
isomorphisms. The initial state of a computation is divided into the
input $A$ and the initial heap $H$, and the final state into the
output $B$ and garbage $G$; using cartesian product ($\times$) in
the classical and tensor product ($\otimes$) in the quantum case. To
actually perform a computation we also need a heap initialisation
constant $h$, which intuitively sets all memory cells in a defined
state, e.g. $0$. In the classical case this is just an element of
the set $h\in H$, while in the quantum case it is an element of the
vector space $h\in\CVec{H}$. Such a computational system can be
visualised by the following diagram:
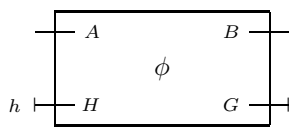
\[
\label{cir:computation}\vcenter{
 \Qcircuit @C=1.5em @R=.5em @!R {
   & \qw & \lstick{A} & &\rstick{B} && \qw\\
   & & & \push{\phi} & &&\\
  \lstick{h} & \eqw & \lstick{H}& &\rstick{G} && \qwe
   \gategroup{1}{2}{3}{6}{1.5em}{-}
 }}
\]

Note that in the above diagram heap inputs are initialised with a
$\vdash$, and garbage outputs are terminated with a $\dashv$. To
summarise, given finite sets $A,B$ a morphism
$(H,h,G,\phi)\in\FCCm{A}{B}$ is given by:
\begin{itemize}
\item a finite set of initial heaps $H$,
\item an initial heap $h \in H$,
\item a finite set of garbage states $G$,
\item a bijection $\phi \in A\times H \simeq B \times G$,
\end{itemize}
while a morphism $(H,h,G,\phi)\in\FQCm{A}{B}$ is given by
\begin{itemize}
\item a finite set $H$, the basis of the space of initial heaps,
\item a heap initialisation vector $h \in \CVec{H}$,
\item a finite set $G$, the basis of the space of garbage states,
\item a unitary operator $\phi \in \UnitaryArr{A\otimes H}{B \otimes G}$.
\end{itemize}
\noindent
Given two computational systems we can compose them by combining
initial and final heaps:

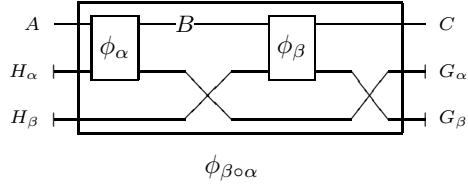
\[
\label{cir:compFCC}\vcenter{
 \Qcircuit @C=1.4em @R=1.2em @!R {
  \lstick{A} & \multigate{1}{\phi_\alpha} &\pushS{B}\qw & \qw & \multigate{1}{\phi_\beta} & \qw &\qw& \rstick{C} \qw\\
  \lstick{H_\alpha} & \ghost{\phi_\alpha}\eqw & \qxT& & \ghost{\phi_\beta} & \qxT && \rstick{G_\alpha} \qwe \\
  \lstick{H_\beta} & \eqw & \qxB &  & \qw & \qxB & & \rstick{G_\beta} \qwe
   \gategroup{1}{2}{3}{7}{1em}{-}\\
  & & & \mbox{$\phi_{\beta\circ \alpha}$} & & &
 }
}
\]
\noindent
More formally, given the morphisms $\alpha$ and $\beta$:
\begin{eqnarray*}
  \alpha &=& (H_\alpha,h_\alpha,G_\alpha,\phi_\alpha) \in \FCCm{A}{B} \\
  \beta  &=& (H_\beta,h_\beta,G_\beta,\phi_\beta) \in \FCCm{B}{C}
\end{eqnarray*}
\noindent
the composite morphism $\beta\circ \alpha = (H,h,G,\phi)$ is given
by:
\begin{eqnarray*}
  H & = & H_\alpha \times H_\beta\\
  h & = & (h_\alpha,h_\beta)\\
  G & = & G_\alpha \times G_\beta\\
  \phi & = & (G_\alpha \times \phi_\beta) \circ (H_\beta \times \phi_\alpha)
\end{eqnarray*}
Note that we have omitted some obvious symmetric monoidal isomorphisms
for $\times$ from the definition of $\phi$. We leave it to the reader
to construct the identity computation.

Analogously, given morphisms
\begin{eqnarray*}
  \alpha &=& (H_\alpha,h_\alpha,G_\alpha,\phi_\alpha) \in \FQCm{A}{B} \\
  \beta  &=& (H_\beta,h_\beta,G_\beta,\phi_\beta) \in \FQCm{B}{C}
\end{eqnarray*}
the composite $\beta\circ \alpha = (H,h,G,\phi)$ is given by
\begin{eqnarray*}
  H & = & H_\alpha \otimes H_\beta\\
  h & = & h_\alpha\otimes h_\beta\\
  G & = & G_\alpha \otimes G_\beta\\
  \phi & = & (G_\alpha \otimes \phi_\beta) \circ (H_\beta \otimes \phi_\alpha)
\end{eqnarray*}
Note that $\otimes$ is actually $\times$ on the underlying finite
sets, since $\CVec{A}\otimes\CVec{B} \simeq \CVec{A\times B}$.
However, we shall use the tensor symbol because we interpret the
constructed set as the basis of the tensor product of the associated
vector spaces.  As in the classical case we omit symmetric monoidal
isomorphisms for $\otimes$.

We consider two computational systems as extensionally equal if they
map the same inputs to the same outputs. That is, for $\FCC$, a
morphism $\alpha=(H,h,G,\phi)\in\FCCm{A}{B}$ gives rise to a
function on finite sets $\forgetC\,\alpha\in A \to B$ by
\begin{equation*}\label{cat:morphismComm1}\vcenter{
  \xymatrix@C+1em@R+1ex{
   A \times H \ar[r]_{\phi}& B \times G \ar[d]^{\pi_1}\\
   A \ar[u]_{(-,h)} \ar[r]_{\forgetC\,\alpha} & B }
 }
\end{equation*}
\noindent
How do we do this for $\FQC$? There is no sensible projection
operation on tensor products. Indeed, forgetting a part of a pure
state (i.e. a vector of the Hilbert space) leads to a mixed state,
which is modelled by a density operator $\delta \in \LinArr{A}{A}$.
This is a positive operator, whose eigenvalues are interpreted
as the probability that the system is in the corresponding
eigenstate. Extensionally, quantum computations give rise to
completely positive mappings, also called superoperators, see
\cite{Hirvensalo}, pp. 136 or \cite{selinger:qpl} for details. Given
$\alpha = (H,h,G,\phi)\in\FQCm{A}{B}$ we write $\adjoin{\phi}\in
\SuperArr{A\otimes H}{B\otimes G}$ for the associated superoperator
$\adjoin{\phi}\,\rho = \phi \circ \rho \circ \adj{\phi}$. The heap
initialisation vector $h\in \CVec{H}$ can be lifted to a density
matrix $\lift{h} \in \Dens{H}$ by $\lift{h}=\ket{h}\bra{h}$.
Combining this with the partial trace operator $\tr_G \in
\SuperArr{B\otimes
  G}{B}$ we obtain $\forgetQ\,\alpha\in \SuperArr{A}{B}$ by
\begin{equation*}\label{cat:morphismComm2}\vcenter{
  \xymatrix@C+1em@R+1ex{
   A \otimes H \ar[r]_{\adjoin{\phi}}& B \otimes G \ar[d]^{\tr_G}\\
   A \ar[u]_{-\otimes\lift{h}} \ar[r]_{\forgetQ\,\alpha} & B }
 }
\end{equation*}
in the category of superoperators.

We say that two computations $\alpha,\beta\in F\,A\,B$ are
extensionally equal ($\alpha \exteq \beta$), if the induced maps are
equal; $\forget{F}\,\alpha = \forget{F}\,\beta$ where
$F\in\{\FCC,\FQC\}$. We define the homsets of $\FCC,\FQC$ as the
quotients of the underlying representation by extensional equality. It
is straightforward to verify that composition respects extensional
equality.


As a consequence of our definition we obtain that the assignment
of maps to computations gives rise to forgetful functors
$\forgetC\in \FCC \to \FinSet$ and $\forgetQ\in \FQC\to\Super$.
Both functors are full
\footnote{In the case of $\FQC$ fullness is a consequence of Kraus'
decomposition theorem.}
and faithful. Hence, our categories $\FCC$ and $\FQC$
can be viewed just as different presentations of $\FinSet$ and
$\Super$. However, going via $\FCC$ and $\FQC$ has the benefit
that we get an implementation of our programs as reversible
circuits in the classical case and quantum circuits in the quantum
case.

An important class of morphisms are the ones which do not produce
garbage, i.e. where $G=1$, they give rise subcategories
$\FCCo,\FQCo$ of \emph{strict} morphisms.
All strict maps are isometries,
i.e. linear maps such that $\braket{f\,v | f\,w} = \braket{v|w}$.
However, not all isometries arise from strict computations.
\footnote{This is only due to dimensional reasons, indeed in the
domain of our interpretation where all spaces are of a size $2^n$
the functor is full.}

While $\FQC$ and $\FCC$ are very similar indeed, the fact that $\FQC$
is based on wave mechanics enables non-local interaction which is
exploited in quantum programming. However, there is also a new
challenge: the possibility of decoherence.  Let $\delta \in 2 \to 2
\times 2$ where $2 = \{0,1\}$ be defined as $\delta\,x = (x,x)$; which
can be easily realised by a \textsc{cnot} gate. The same
implementation gives rise to $\LinArr{\Qbit}{\Qbit \otimes \Qbit}$,
writing $\Qbit$ for the object $2$ in $\FQC$. In either case, we
can compose this with $\pi_1 \in 2\times 2\to 2$ (or $\pi_1\in
\SuperArr{\Qbit\otimes\Qbit}{\Qbit}$) which leads to the following
picture:
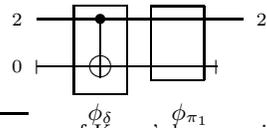
\[
\vcenter{

\Qcircuit@C=1em @R=1em @!R{
 \lstick{2} & \qw & \ctrl{1} & \qw & \qw &\qw  &\qw&\rstick{2}\qw\\
 \lstick{0} & \eqw & \targ    & \qw & \qw &\qw &\qwe \\
  &&\mbox{\small{$\phi_\delta$}}&&&\mbox{\small{$\phi_{\pi_1}$}}\gategroup{1}{3}{2}{3}{1.2em}{-}\gategroup{1}{5}{2}{6}{1em}{-}
}}
\]
Clearly, classically we have just defined an inefficient version of
the identity $\pi_1 \circ \delta = \Id$; we copy a bit and then
throw the copy away. However, the situation is quite different in
the quantum case: while the implementation is given by the same
diagram by replacing classical reversible circuits with quantum
circuits, the composition is not the identity, it is a
\emph{measurement} operation. That is, if we input a pure state like
$R = \{{1\over\sqrt{2}}\ket{0} + {1\over\sqrt{2}}\ket{1}\}$ the
output is a mixed state ${1\over 2}\{\ket{0}\}+{1\over
2}\{\ket{1}\}$ corresponding to a random qubit. We have lost the
advantage of quantum computation and are back in the world of
classical probabilistic computations.

As a consequence of this observation we draw the conclusion that one
of the main issues a quantum programming language has to address is
the \emph{control of decoherence}.  This is somehow the opposite of
the common view which insists that the \emph{no cloning theorem}
outlaws contraction. We observe that the implementation of $\delta$
shares a qubit, but it doesn't clone it; considering $R$ again we
obtain the EPR state $\{{1\over\sqrt{2}}\ket{00} +
{1\over\sqrt{2}}\ket{11}\}$
 after executing only $\delta$. We claim
that this is a natural explanation of contraction because it is
completely uniform in both the classical and the quantum case.
Indeed, classical functional languages do not implement contraction
by copying data either.
$\delta$ is strict and therefor maps pure states to pure states.
In contrast, operations like $\pi_1$ are interpreted
by a non-trivial partial trace which introduces decoherence. Hence
it is \emph{weakening} which deserves our attention, not
\emph{contraction}.

\section{QML: Rules and semantics}
\label{sec:qml}

We introduce here the typing rules and the denotational semantics of
QML, the latter gives rise to a compilation of QML programs to quantum
circuits. The compilation is presented diagramatically, implementing
it requires some care to make sure that the wires generated by
subcomputations match as intended.

\subsection{Typing rules}

We will only present the typed syntax of QML, which is based on
strict linear logic, the untyped syntax is implicit in the typed one.
We do allow explicit weakenings annotating a
term by by a context. This leads to an unambiguous type assignment.
Any weakening will be translated into the use of a non-trivial
partial trace, and hence decoherence in the \emph{denotational
semantics}.  Another source of
decoherence is the use of \emph{case}, or its special instance
\emph{if-then-else}. We make this explicit by introducing two
different case-operators: one which observes a qubit and thus leads
to decoherence; and another which is free of decoherence but
requires that we derive that the two alternatives live in orthogonal
spaces. For this purpose we introduce a judgement $t \perp u$.
Another novelty of our language is a \emph{term--former} to create
superpositions; we can,for example, write \ensuremath{\{\mskip1.5mu (\Cnid{qtrue},\Cnid{qtrue})\;\mid\;(\Cnid{qfalse},\Cnid{qfalse})\mskip1.5mu\}}, to create an EPR state.  Note that we are
ignoring the factor $1\over\sqrt{2}$ which can be automatically
inserted by the compiler. The construction of a superposition also
requires to show that the participating terms are orthogonal.

Our basic typing judgements are $\Gamma\vdash t:\sigma$ meaning
that $t$ has type $\sigma$ under context $\Gamma$.
and $\Gamma\vdasho t:\sigma$ for strict terms. We embed
$\vdasho$ in $\vdash$:
\[\ru{\Gamma\vdasho t:\sigma}
   {\Gamma\vdash t:\sigma}
\]
To avoid repetition, we also use the schematic judgements $\Gamma\vdash^a
t:\sigma$ where $a\in\{-,\circ\}$. We use $\sigma$,$\tau$ and $\rho$
to quantify over types, which are generated by $1,\sigma \oplus
\tau,\sigma \otimes \tau$. Qubits are defined as \ensuremath{\Tyid{Q_2}\mathrel{=}\mathrm{1}\oplus\mathrm{1}}.

$\Gamma$ is a context, i.e. a function from a finite set of
variables $\dom{\G}$ into the set of types. We write contexts as
$\Gamma = x_1:\tau_1,\dots,x_n:\tau_n$ and use $\emptyC$ for the
empty context.  $\Gamma,x:\tau$ is the context $\G$ extended by
$x:\tau$. This operation is only defined if $\Gamma$ does not
already assign a type to $x$.

For the additive rules, we introduce the operator $\otimes$ mapping pairs of
contexts to contexts:
\[\begin{array}{lcll}
  \Gamma,x:\sigma \otimes \Delta,x:\sigma & = &
  (\Gamma\otimes\Delta),x:\sigma\\
  \Gamma,x:\sigma \otimes \Delta & = & (\Gamma \otimes
  \Delta),x:\sigma & \mbox{if $x \notin \dom{\Delta}$}\\
  \emptyC \otimes \Delta & = & \Delta
\end{array}\]
\noindent
This operation is partial -- it is only well-defined if the two
contexts do not assign different types to the same variable.

\subsection{Denotational semantics}

We assign to every type $\sigma$ the number $|\sigma|$ which is the
size of a quantum register needed to store elements of $\sigma$, we also
interpret expressions of the form $\sigma\sqcup\tau$:
\begin{eqnarray*}
  |1| & = & 0 \\
  |\sigma\sqcup \tau| & = & \max\; \{|\sigma|,|\tau|\}\\
  |\sigma \oplus \tau| & = & |\sigma\sqcup \tau|+1\\
  |\sigma \otimes \tau| & = & |\sigma|+|\tau|
\end{eqnarray*}

The interpretation of a type is the $\FQC$ object of quantum registers of the
right size: $\den{\sigma} = \Qbit^{|\sigma|}$.
Contexts $\Gamma = x_1:\tau_1,\dots,x_n:\tau_n$ are interpreted as
the tensor product of their components
$\den{\G} = \den{\tau_1}\ot\den{\tau_2}\ot\dots\ot\den{\tau_n}$.
A typing derivation $\G\vdash t:\sigma$ is interpreted as an $\FQC$ morphism
$\den{t} \in \FQCm{\den{\G}}{\den{\sigma}}$, correspondingly, $\G\vdasho t:\sigma$
is interpreted as $\den{t} \in \FQCom{\den{\G}}{\den{\sigma}}$.

The interpretation of orthogonality is more involved. Given $\Gamma
\vdasho t:\sigma$ and $\Gamma' \vdasho u:\sigma$ where $|\Gamma| =
|\Gamma'|$ we interpret a derivation $t\perp u$ as a structure
$(S,f,g,\psi)$ where $S$ is an object of $\FQC$, $l\in
\FQCm{\den{\G}}{S}$, $g\in\FQCm{\den{\G'}}{S}$ such that $\den{t} =
\phi\circ(\qtrue\ot -)\circ f$ and $\den{u} = \phi\circ(\qfalse\ot
-)\circ g$.

To interpret the operator $\ot$ on contexts we define an $\FQCo$ morphism
$\CO_{\G,\D}\in\FQCom{\den{\G\ot\D}}{(\den{\G}\ot\den{\D})}$
\[
\Qcircuit@C=1em @R=1em @!R{
\lstick{\Gamma\ot\D}&\Cgg&\rstick{\Gamma}\qw\\
\lstick{H_{\Gamma,\D}}&\gCgg\eqw&\rstick{\D}\qw\\
}
\]
by induction over the definition of $\G\ot\D$: If a variable
$x:\sigma$ appears in both contexts we have to use
$\delta_\sigma\in\FQCom{\den{\sigma}}{(\den{\sigma}\ot\den{\sigma})}$
which generalises $\delta_2$, discussed earlier, by applying it in
parallel to all qubits. All the other cases can be dealt with by
applying monoidal isomorphisms. Similarly, we define an explicit
weakening operator $\WK_{\G,\D}\in\FQCm{\den{\G\ot\D}}{\den{\G}}$.

\subsection{Structural rules}
\label{sec:struct}

We start with the strict variable rule and the non-strict weakening
and their interpretations
\[\begin{array}{cc}
  \Ax{x \tin \sigma\vdasho x \tin \sigma}{var}
  & \qquad
    \Ru
    {\G \vdash t \tin \sigma}
    {\G \ot \D \vdash t^{\dom{\Delta}} \tin \sigma}
    {weak}\\
\Qcircuit@C=1.2em @R=1.2em @!R{
 \lstick{\sigma}&\qw&\rstick{\sigma}\qw\\
}
&\qquad
\Qcircuit@C=0.7em @R=0.7em @!R{
 \lstick{\G \ot \D}&\multigate{1}{\phi_{\WK_{\Gamma,\Delta}}}&\pushT{\G}\qw&\qw&\multigate{1}{\phi_t}&\rstick{\sigma}\qw \\
                   &\ghostX{\phi_{\WK_{\Gamma,\Delta}}}      &\qxT         &   &\ghost{\phi_t}       &\rstick{G_t}\qwe\\
 \lstick{H_t}      &\eqw                                     &\qxB         &   &\qw                  &\rstick{G_{\G-\D}}\qwe\\
}
\\
\end{array}
\]
\noindent
Next, we introduce a let-rule which is also the basic vehicle to define
first order programs.
\[
  \Ru{\begin{array}{l}
      \G \vdash^a t \tin \sigma \\
      \D,\, x \tin \sigma \vdash^b u \tin \tau\\
    \end{array}}
     {\G\ot\D \vdash^{a\sqcap b} \rlet\, x = t \,\rin\, u \tin \tau}
     {let}
\]
$\circ \sqcap \circ = \circ$ and $-$ otherwise. We leave the condition
that $\Gamma \otimes \Delta$ is defined as an implicit precondition of
this and subsequent rules using $\otimes$. The interpretation of the
let-rule is given by the following circuit:
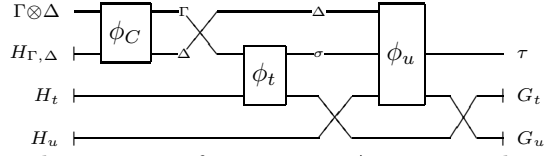
\[
\Qcircuit@C=1em @R=1em @!R{
 \lstick{\Gamma\ot\D}  &\Cgg\qw  &\lab{\Gamma}\qxT &   &\qw                  &\lab{\D}\qw     &\qw&\multigate{2}{\phi_u}&    &\\
 \lstick{H_{\Gamma,\D}}&\gCgg\eqw&\lab{\D}\qxB     &   &\multigate{1}{\phi_t}&\lab{\sigma}\qw &\qw&\ghost{\phi_u}       &\qw &\qw&\rstick{\tau}\qw\\
 \lstick{H_t}          &\eqw     &\qw              &\qw&\ghost{\phi_t}       &\qxT            &   &\ghost{\phi_u}       &\qxT&   &\rstick{G_t}\qwe\\
 \lstick{H_u}          &\eqw     &\qw              &\qw&\qw                  &\qxB            &   &\qw                  &\qxB&   &\rstick{G_u}\qwe\\
}
\]

Weakenings can affect the meaning of a program. As an example
consider:
\[ y:\Qbit \vdash \rlet\, x = y \,\rin\, x^{\{\}} : \Qbit \]
This program will be interpreted as the identity circuit, in
particular it is decoherence-free. However, consider
\[ y:\Qbit \vdash \rlet\, x = y \,\rin\, x^{\{y\}} : \Qbit \]
This program is interpreted by a circuit equivalent to the one
corresponding to $\pi_1 \circ \delta$ shown earlier; hence it
introduces a measurement.

\subsection{Rules for $\otimes$}
\label{sec:otimes}

The rules for $1$, $\otimes$ are the standard rules from linear logic. In the case of $1$
instead of an explicit elimination rule we allow implicit weakening:
\[
  \Ax{\emptyC \vdasho ()\tin 1}
  {1-intro}
  \qquad
  \Ru{\G,x\tin 1 \vdash^a t : \sigma}
     {\G\vdash^a t:\sigma}
     {1-weak}
\]
The interpretation of the rules for $1$ in terms of circuits is invisible, since
$1$ doesn't carry any information. The interpretation of the rules for $\otimes$
is more interesting --- the introduction rule simply merges the components

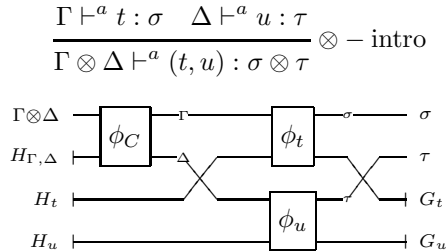
\[\begin{array}{c}
\Ru{\G \vdash^a t \tin \sigma \quad \D \vdash^a u \tin \tau}
   {\G \ot \D \vdash^a (t,u) \tin \sigma \ot \tau}
   {\ot-intro}
\\
\Qcircuit@C=1em @R=1em @!R{
 \lstick{\G \ot \D}&\Cgg     &\lab{\G}\qw &\qw&\qw&\multigate{1}{\phi_t}&\lab{\sigma}\qw&\qw&\rstick{\sigma}\qw\\
 \lstick{H_{\G,\D}}&\gCgg\eqw&\lab{\D}\qxT&   &\qw&\ghost{\phi_t}\qw    &\qxT           &   &\rstick{\tau}\qw\\
 \lstick{H_t}      &\eqw     &\qxB        &   &\qw&\multigate{1}{\phi_u}&\lab{\tau}\qxB &   &\rstick{G_t}\qwe\\
 \lstick{H_u}      &\eqw     &\qw         &\qw&\qw&\ghost{\phi_u}\qw    &\qw            &\qw&\rstick{G_u}\qwe
}
\end{array}
\]
The interpretation of the elimination rule is similar to the let-rule:
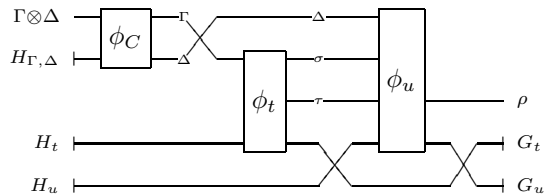
\[
  \Ru{\begin{array}{l}
      \G \vdash^a t \tin \sigma \ot \tau\\
      \D,\, x \tin \sigma, y \tin \tau \vdash^b u \tin \rho\\
    \end{array}}
     {\G\ot\D \vdash^{a\sqcap b} \rlet (x,y) = t \rin u \tin \rho}
     {\ot-elim}
\]
\[
\Qcircuit@C=1em @R=1em @!R{
 \lstick{\G\ot\D}  &\Cgg\qw  &\lab{\Gamma}\qxT &   &\qw                  &\lab{\D}\qw    &\qw&\multigate{3}{\phi_u}&   &\\
 \lstick{H_{\G,\D}}&\gCgg\eqw&\lab{\D}\qxB     &   &\multigate{2}{\phi_t}&\lab{\sigma}\qw&\qw&\ghost{\phi_u}&      &   &\\
                   &         &                 &   &\ghostX{\phi_t}      &\lab{\tau}\qw  &\qw&\ghost{\phi_u}&\qw   &\qw&\rstick{\rho}\qw\\
 \lstick{H_t}      &\eqw     &\qw              &\qw&\ghost{\phi_t}       &\qxT           &   &\ghost{\phi_u}&\qxT  &   &\rstick{G_t}\qwe\\
 \lstick{H_u}      &\eqw     &\qw              &\qw&\qw                  &\qxB           &   &\qw           &\qxB  &   &\rstick{G_u}\qwe\\
}
\]
As an example, here is a simple program which swaps two qubits:
\[ p:\Qbit\otimes\Qbit \vdash \rlet\,(x,y) = p \rin (y^{\{\}},x^{\{\}}) : \Qbit\otimes\Qbit \]
Again it is important to mark the variables with the empty set of
variables. The alternative program
\[ p:\Qbit\otimes\Qbit \vdash \rlet\,(x,y) = p \rin (y^{\{p\}},x^{\{p\}}) : \Qbit\otimes\Qbit
\]
would measure the qubits while swapping them.

\subsection{Rules for $\oplus$}
\label{sec:oplus}

We represent values in $\sigma\oplus\tau$ as words of fixed length,
as in classical computing. Unfolding our type interpretation we have
that $\den{\sigma\oplus\tau} = \Qbit\ot\den{\sigma\sqcup\tau}$ where
$\den{\sigma\sqcup\tau}$ can store a value either of $\den{\sigma}$
or $\den{\tau}$. To adjust the size we use an easily definable
padding operator
$\PAD_{\sigma\sqcup\tau}\in\FQCm{\den{\sigma}}{\den{\sigma\sqcup\tau}}$,
which simply sets unused bits to $0$.

The introduction rules for $\oplus$ are the usual classical rules
for $+$; note that they preserve strictness.
\[\begin{array}{c}
  \Ru{\Gamma \vdash^a s \tin \sigma}
  {\Gamma \vdash^a \rinl\, s \tin \sigma \oplus \tau}
  {+ \,\mathrm{intro}_1}
  \\
  \\                              
  \Qcircuit@C=1em @R=1em {
    \lstick{\G}     &\multigate{1}{\phi_s}&\lab{\sigma}\qw &\qw&\multigate{1}{\phi_{\PAD_{\sigma\sqcup\tau}}}\\
    \lstick{H_s}    &\ghost{\phi_s}\eqw   &\qxT            &   &\ghost{\phi_{\PAD_{\sigma\sqcup\tau}}}\qw&\qw &\qw&\rstick{\sigma\sqcup \tau}\qw\\
    \lstick{H_{t-s}}&\eqw                 &\qxB            &   &\qw                                      &\qxT&   &\rstick{\Qbit}\qw\\
    \lstick{\Qbit}  &\gate{X}eqw          &\qw             &\qw&\qw                                      &\qxB&   &\rstick{G_s}\qwe\\
  }\\
  \\
  \Ru{\Gamma \vdash^a t \tin \tau}
  {\Gamma \vdash^a \rinr\, t \tin \sigma \oplus \tau}
  {+ \,\mathrm{intro}_2}\\
  \Qcircuit@C=1em @R=1em @!R{
    \lstick{\G}     &\multigate{1}{\phi_t}&\lab{\tau}\qw &\qw&\multigate{1}{\phi_{\PAD_{\tau\sqcup\sigma}}}\\
    \lstick{H_t}    &\ghost{\phi_t}\eqw   &\qxT          &   &\ghost{\phi_{\PAD_{\tau\sqcup\sigma}}}&\qw &\qw&\rstick{\sigma\sqcup \tau}\qw\\
    \lstick{H_{t-s}}&\eqw                 &\qxB          &   &\qw                                   &\qxT&   &\rstick{\Qbit}\qw\\
    \lstick{\Qbit}  &\eqw                 &\qw       &\qw&\qw                                   &\qxB&   &\rstick{G_t}\qwe\\
  }
\end{array}
\]
\noindent
where $X$ is negation.

We define $\qtrue^X = \rinl\,()^X : \Qbit$ and $\qfalse^X =
\rinr\,()^X : \Qbit$.  To be able to interpret case expressions we
introduce a biconditional operation on unitary operators. Given
$\phi,\psi \in \UnitaryArr{A}{B}$ we construct
\[\choice{\phi}{\psi}\in \UnitaryArr{\Qbit\ot A}{\Qbit\ot B}\] by the following matrix
\[\begin{array}{llcl}
  \choice{\phi}{\psi}\,& (\true,a)\,(\true,b) & = & \phi\,a\,b\\
  \choice{\phi}{\psi}\,& (\false,a)\,(\false,b) & = & \psi\,a\,b\\
  \choice{\phi}{\psi}\,& (x,a)\,(y,b) & = & 0\quad\mbox{everywhere else}
\end{array}\]
As already indicated we have two different elimination rules --- we
begin with the one which measures a qubit, since it is basically the
classical rule modulo additivity of contexts.
\[
  \Ru{\begin{array}{l}
       \G \vdash c \tin \sigma\oplus \tau\\
       \D,\, x \tin \sigma \vdash t \tin \rho\\
       \D,\, y \tin \tau \vdash u \tin \rho\\
      \end{array}}
     {\G\ot\D \vdash \rcase\ c \rof \{\rinl x \Rightarrow t\ \vert\ \rinr y\Rightarrow u \} \tin \rho}
     {\oplus -elim}
\]

We have $\den{t}\in\FQCm{\den{\Delta\ot\sigma}}{\den{\rho}}$ and
$\den{u}\in\FQCm{\den{\Delta\ot\tau}}{\den{\rho}}$. By padding the
input we turn them into
$\ceil{\den{t}},\ceil{\den{u}}\in\FQCm{\den{\Delta\ot(\sigma\sqcup\tau)}}{\den{\rho}}$.
There is no reason why the size of the associated heap and garbage
should be the same, however, we have that $H_t + G_u = H_u + G_t$ and hence
we can stretch both maps uniformly to $H = H_t \sqcup H_u$ and
$G = G_t \sqcup G_u$ giving rise to $\phi_{\ceil{\den{t}}}$ and
$\phi_{\ceil{\den{u}}}$ of identical dimensions. Hence
we can apply the choice operator to construct 
$\psi = \choice{\phi_{\ceil{\den{t}}}}{\phi_{\ceil{\den{u}}}}$,
and with some plumbing we obtain:

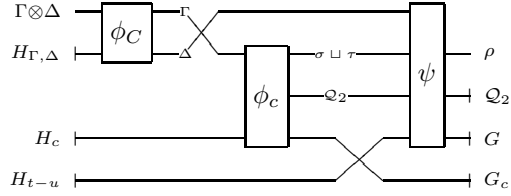
\[
 \Qcircuit@C=1em @R=1em @!R{
  \lstick{\G\ot\D}  &\Cgg     &\lab{\G}\qxT&   &\qw                  &\qw           &\qw&\multigate{3}{\psi}&\\
  \lstick{H_{\G,\D}}&\gCgg\eqw&\lab{\D}\qxB&   &\multigate{2}{\phi_c}&\lab{\sut}\qw &\qw&\ghost{\psi}       &\rstick{\rho}\qw\\
                    &         &            &   &\ghostX{\phi_c}      &\lab{\Qbit}\qw&\qw&\ghost{\psi}       &\rstick{\Qbit}\qwe\\
  \lstick{H_c}      &\eqw     &\qw         &\qw&\ghost{\phi_c}       &\qxT          &   &\ghost{\psi}       &\rstick{G}\qwe\\
  \lstick{H_{t-u}}  &\eqw     &\qw         &\qw&\qw                  &\qxB          &   &\qw                                   &\rstick{G_c}\qwe\\
 }
\]

\noindent
We can derive if-then-else as

\begingroup\par\noindent\advance\leftskip\mathindent\(
\begin{pboxed}\SaveRestoreHook
\column{B}{@{}l@{}}
\column{3}{@{}l@{}}
\column{8}{@{}l@{}}
\column{E}{@{}l@{}}
\fromto{3}{E}{{}\mathbf{if}\;\Varid{b}\;\mathbf{then}\;\Varid{t}\;\mathbf{else}\;\Varid{u}\mathrel{=}{}}
\nextline
\fromto{3}{8}{{}\hsindent{5}{}}
\fromto{8}{E}{{}\mathbf{case}\;\Varid{b}\;\mathbf{of}\;\{\mskip1.5mu \Varid{inl}\;\anonymous \Rightarrow \Varid{t}\mid \Varid{inr}\;\anonymous \Rightarrow \Varid{u}\mskip1.5mu\}{}}
\ColumnHook
\end{pboxed}
\)\par\noindent\endgroup\resethooks

\noindent
and use this to implement a form of negation:

\begingroup\par\noindent\advance\leftskip\mathindent\(
\begin{pboxed}\SaveRestoreHook
\column{B}{@{}l@{}}
\column{3}{@{}l@{}}
\column{E}{@{}l@{}}
\fromto{3}{E}{{}\Varid{mnot}\mathbin{:}\Tyid{Q_2}\multimap\Tyid{Q_2}{}}
\nextline
\fromto{3}{E}{{}\Varid{mnot}\;\Varid{x}\mathrel{=}\mathbf{if}\;\Varid{x}\;\mathbf{then}\;\Cnid{qfalse}\;\mathbf{else}\;\Cnid{qtrue}{}}
\ColumnHook
\end{pboxed}
\)\par\noindent\endgroup\resethooks

\noindent
However, this program will measure the qubit before negating it. If we
want to avoid this we have to use the decoherence-free version of
case, which relies on the orthogonality judgement: $ t \perp u $,
which is defined for terms in the same type and context $\G\vdash
t,u:A$. We will introduce the rules for orthogonality later.
Intuitively, $t\perp u$ holds if the outputs $t$ and $u$ are always
orthogonal, e.g. we will be able to derive
$\qtrue^{\{\}}\perp\qfalse^{\{\}}$. Hence, we introduce the strict
case by:

\[
  \Ru{\begin{array}{lr}
       \G \vdash^a c \tin \sigma\oplus \tau\\
       \D,\ x \tin \sigma \vdasho t \tin \rho\\
       \D,\ y \tin \tau   \vdasho u \tin \rho \quad& t \perp u\\
      \end{array}}
     {\begin{array}{lcl}
       \G\ot\D &\vdash^a & \rcaseo \ c \rof \\
               &         & \{\rinl x \Rightarrow t\ \vert\ \rinr y \Rightarrow u \} \tin \rho\\
     \end{array}}
     {\oplus -elim^\circ}
\]

\noindent
It turns out that there is no sensible way to define $\rcaseo$ if
$\sigma$ and $\tau$ have different sizes. Hence we define the orthogonality
judgement in a way that it only succeeds, if $|\sigma| = |\tau|$
and hence $\den{\sigma}=\den{\tau}$.

To define the interpretation, we have to exploit the data from the
orthogonality judgement $\den{t\perp u} = (S,f,g,\psi)$ where
$\psi\in\UnitaryArr{S \ot \Qbit}{\den{\rho}}$ and
$f,g\in\FQCom{(\den{\Delta}\ot\den{\sigma})}{S}$. We note that both
morphisms must have the same heap and hence we can construct
\[ \choice{\phi_f}{\phi_g} \in \FQCom{(\Qbit\ot\den{\Delta}\ot\den{\sigma})}{(\Qbit\ot S)}.\]
Now, the main observation is that we just have to apply the unitary
operator $\phi_{t \perp u}$ to make the qubit disappear, leading to
the following diagram:
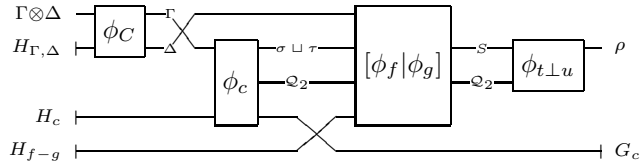
\[
 \Qcircuit@C=0.7em @R=0.7em @!R{
  \lstick{\G\ot\D}  &\Cgg     &\lab{\G}\qxT&   &\qw                  &\qw           &\qw&\multigate{3}{\choice{\phi_f}{\phi_g}}&\\
  \lstick{H_{\G,\D}}&\gCgg\eqw&\lab{\D}\qxB&   &\multigate{2}{\phi_c}&\lab{\sut}\qw &\qw&\ghost{\choice{\phi_f}{\phi_g}}       &\lab{S}\qw&\multigate{1}{\phi_{t\perp u}}&\rstick{\rho}\qw\\
                    &         &            &   &\ghostX{\phi_c}      &\lab{\Qbit}\qw&\qw&\ghost{\choice{\phi_f}{\phi_g}}       &\lab{\Qbit}\qw &\ghost{\phi_{t\perp u}}\\
  \lstick{H_c}      &\eqw     &\qw         &\qw&\ghost{\phi_c}       &\qxT          &   &\ghost{\choice{\phi_f}{\phi_g}}       \\
  \lstick{H_{f-g}}  &\eqw     &\qw         &\qw&\qw                  &\qxB          &   &\qw &\qw&\qw                        &\rstick{G_c}\qwe\\
 }
\]

\noindent
Note that we only allow strict terms in the branches of a strict case. In a previous
draft of this paper we tried to be more liberal, however, this causes problems because
the qubit we are branching over can be indirectly measured by the garbage. This problem
was pointed out by Peter Selinger.

Using the decoherence-free version \ensuremath{\mathbf{if}^\circ} we can implement standard reversible and
hence quantum operations such as \ensuremath{\Varid{qnot}}:
\begingroup\par\noindent\advance\leftskip\mathindent\(
\begin{pboxed}\SaveRestoreHook
\column{B}{@{}l@{}}
\column{3}{@{}l@{}}
\column{13}{@{}l@{}}
\column{E}{@{}l@{}}
\fromto{3}{E}{{}\Varid{qnot}\mathbin{:}\Tyid{Q_2}\multimap\Tyid{Q_2}{}}
\nextline[\blanklineskip]
\fromto{3}{13}{{}\Varid{qnot}\;\Varid{x}\mathrel{=}{}}
\fromto{13}{E}{{}\mathbf{if}^\circ\;\Varid{x}{}}
\nextline
\fromto{13}{E}{{}\mathbf{then}\;\Cnid{qfalse}{}}
\nextline
\fromto{13}{E}{{}\mathbf{else}\;\Cnid{qtrue}{}}
\ColumnHook
\end{pboxed}
\)\par\noindent\endgroup\resethooks

\noindent
and the conditional not \ensuremath{\Varid{cnot}}:

\begingroup\par\noindent\advance\leftskip\mathindent\(
\begin{pboxed}\SaveRestoreHook
\column{B}{@{}l@{}}
\column{3}{@{}l@{}}
\column{15}{@{}l@{}}
\column{30}{@{}l@{}}
\column{E}{@{}l@{}}
\fromto{3}{E}{{}\Varid{cnot}\mathbin{:}\Tyid{Q_2}\multimap\Tyid{Q_2}\multimap\Tyid{Q_2}\otimes\Tyid{Q_2}{}}
\nextline[\blanklineskip]
\fromto{3}{15}{{}\Varid{cnot}\;\Varid{c}\;\Varid{x}\mathrel{=}{}}
\fromto{15}{E}{{}\mathbf{if}^\circ\;\Varid{c}{}}
\nextline
\fromto{15}{30}{{}\mathbf{then}\;(\Cnid{qtrue},{}}
\fromto{30}{E}{{}\Varid{qnot}\;\Varid{x}){}}
\nextline
\fromto{15}{30}{{}\mathbf{else}\;(\Cnid{qfalse},{}}
\fromto{30}{E}{{}\Varid{x}){}}
\ColumnHook
\end{pboxed}
\)\par\noindent\endgroup\resethooks

\noindent
and finally the Toffolli operator which is basically a conditional \ensuremath{\Varid{cnot}}:

\begingroup\par\noindent\advance\leftskip\mathindent\(
\begin{pboxed}\SaveRestoreHook
\column{B}{@{}l@{}}
\column{3}{@{}l@{}}
\column{17}{@{}l@{}}
\column{23}{@{}l@{}}
\column{33}{@{}l@{}}
\column{E}{@{}l@{}}
\fromto{3}{E}{{}\Varid{toff}\mathbin{:}\Tyid{Q_2}\multimap\Tyid{Q_2}\multimap\Tyid{Q_2}\multimap\Tyid{Q_2}\otimes(\Tyid{Q_2}\otimes\Tyid{Q_2}){}}
\nextline[\blanklineskip]
\fromto{3}{17}{{}\Varid{toff}\;\Varid{c}\;\Varid{x}\;\Varid{y}\mathrel{=}{}}
\fromto{17}{E}{{}\mathbf{if}^\circ\;\Varid{c}{}}
\nextline
\fromto{17}{23}{{}\mathbf{then}\;{}}
\fromto{23}{33}{{}(\Cnid{qtrue},{}}
\fromto{33}{E}{{}\Varid{cnot}\;\Varid{x}\;\Varid{y}){}}
\nextline
\fromto{17}{23}{{}\mathbf{else}\;{}}
\fromto{23}{33}{{}(\Cnid{qfalse},{}}
\fromto{33}{E}{{}(\Varid{x},\Varid{y})){}}
\ColumnHook
\end{pboxed}
\)\par\noindent\endgroup\resethooks


\subsection{Superpositions}
\label{sec:super}

There is a simple syntactic translation we use to reduce the
superposition operator to the problem of creating an arbitrary
1-qubit state:
\[
  \ru{\begin{array}{ll}
    \Gamma \vdasho t,u \tin \sigma & t \perp u \\
    \abs{\abs{\lambda}}^2+\abs{\abs{\lambda^\prime}}^2 = 1 & \lambda,\lambda^\prime \neq 0 \\
  \end{array}}
  {\begin{array}{lcl}
   \G &\vdasho & \{(\lambda)t \,\vert\, (\lambda^\prime)u\} \tin \sigma\\
      & \equiv & \rifo \{(\lambda)\qtrue \,\vert\, (\lambda^\prime)\qfalse\}\\
      &        & \rthen t \relse u
      \end{array}}
\]

\noindent
The algorithm for the preparation of the one-qubit state to a given degree of precision
(which is a parameter of the compilation) can be obtained from the one-qubit case of
the Kitaev-Solovay theorem, see \cite{NC00}, page 616-624.

\subsection{Orthogonality}
\label{sec:orth}

Given $\Gamma\vdash t:\sigma$ and $\Delta\vdash u:\sigma$ where
$|\Delta|=|\Gamma|$ we define $t\perp u$ by the following rules.
The idea of $t\perp u$ is that there is a boolean observation which
tells the two terms apart in every environment. The interpretation
$\den{t\perp u}=(S,f,g,\psi)$ is defined by induction over the
derivations. We present here a sound but incomplete formalisation
of orthogonality, achieving completeness is subject of further work.

\[
\ru{\G\vdasho t : \sigma\qquad \G\vdasho u:\tau}
{\rinl t \perp \rinr u \qquad \rinr t \perp \rinl u}
\]
Here $\rho = \sigma \oplus \tau$, we set $S = \sigma \sqcup \tau$.
In both cases $f$ is obtained by
interpreting $t$ combined with padding and $r$ is given by the
interpretation of $u$ and padding. The circuits for $\psi$ for these
rules are given by:
\[
 \Qcircuit@C=2em @R=1.75em {
 \lstick{S}    &\qw&\\
 \lstick{\Qbit}&\qw&\ustick{_\rho}
 \gategroup{1}{2}{2}{2}{1em}{\}}}
\quad \quad \quad
 \Qcircuit@C=1.5em @R=.75em @!R{
 \lstick{S}    &\qw     &\qw&\\
 \lstick{\Qbit}&\gate{X}&\qw&\urstick{_\rho}
 \gategroup{1}{3}{2}{3}{1em}{\}}}
\]

\[
\ru{t\perp u}{\rinl t\perp\rinl u \quad \rinr t\perp\rinr u}
\]
Let $\Gamma \vdasho \rinl t,\rinl u : \sigma \oplus \tau$ and let
$(S,f,g,\psi)$ be the interpretation of $t\perp u$. From this data we
are constructing the interpretation of $\rinl t\perp\rinl u$ as
$(S,f',g',\psi')$.  We set $S' = S\otimes Q_2\otimes H$ where
$H$ is the heap needed by \ensuremath{\Varid{inl}}. We construct
$f'$ and $g'$ by applying \ensuremath{\Varid{inl}} to $l,r$ on the level of
semantics using the appropriate part of $S'$ as the heap.
$\psi$ is given by the following diagram:
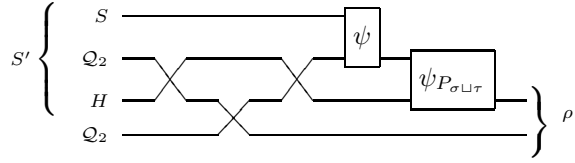
\newcommand{\fudge}{*!R!<1em,0em>=<0em>{_{S'}}}

\[
 \Qcircuit@C=1.2em @R=1em {
       &&\lstick{S}               &\qw &\qw&\qw &\qw&\qw &\qw&\multigate{1}{\psi}\\
 \fudge&&\lstick{\Qbit}           &\qxT&   &\qw &\qw&\qxT&   &\ghost{\psi}&\multigate{1}{\psi_{P_{\sigma\sqcup\tau}}}\\
       &&\lstick{H}&\qxB&   &\qxT&   &\qxB&   &\qw                &\ghost{\psi_{P_{\sigma\sqcup\tau}}}&\qw&\\
       &&\lstick{\Qbit}           &\qw &\qw&\qxB&   &\qw &\qw&\qw                &\qw                                &\qw&\urstick{_\rho}
\gategroup{3}{12}{4}{12}{1em}{\}} \gategroup{1}{1}{3}{1}{1em}{\{}
 }
\]
The second rule for \ensuremath{\Varid{inr}} is done symmetrically.

\[
\ru{t\perp u}{(t,v)\perp (u,w)\quad (v,t)\perp (w,u)}
\]
As above, let $\Gamma \vdasho (t,v),(u,w) : \sigma \otimes \tau$ and
let $(S,f,g,\psi)$ be the interpretation of $t\perp u$
to construct the interpretation of $(t,v)\perp (u,w)$ as $(S',f',g',\psi')$.
We set $S'= S \otimes \tau$ and construct $f'$ and $g'$
by pairing with \ensuremath{\Varid{v}},\ensuremath{\Varid{w}}, semantically.

The definition of $\psi'$ is given by the following diagram:
\[
 \Qcircuit@C=1em @R=1.2em {
            && S &&\qw &\qw&\multigate{1}{\psi}&                  &&\\
 \ulstick{S'}&&{\tau}    &&\qxT&   &\ghost{\psi}       &\rstick{\sigma}\qw&&\\
            &&{Q_2}   &&\qxB&   &\qw                       &\rstick{\tau}\qw  &&\urstick{_\rho}
\gategroup{1}{2}{2}{2}{1em}{\{}\gategroup{2}{9}{3}{9}{1em}{\}}
}
\]

\[ \ru{t \perp u \quad \lambda_0^*\kappa_0 = - \lambda_1^*\kappa_1}
      {\{(\lambda_0)t\ \vert\ (\lambda_1)u\}\perp\{(\kappa_0)t\ \vert\ (\kappa_1)u\}}
\]
As before, assume as given the interpretation of $t \perp u$ as $(S,f,g,\psi)$.
We construct the interpretation of the conclusion as $(S,f,g,\psi')$ where $\psi'$
is given as
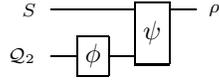
\[\Qcircuit@C=1em @R=0.7em {
  \lstick{S}    &\qw           &\multigate{1}{\psi}\qw&\rstick{\rho}\qw\\
  \lstick{\Qbit}&\gate{\phi}\qw&\ghost{\psi}\qw       &
  }
\]
\noindent
using the rotation $\phi \in \UnitaryArr{\Qbit}{\Qbit}$ given by 
\[\phi = \begin{pmatrix}
  \lambda_0 & \lambda_1 \\
  \kappa_0 & \kappa_1 \\
  \end{pmatrix}
\]

\subsection{Programs}
\label{sec:prog}

So far we have introduced a language of expressions. It is
straightforward to extend this to a notion of first order programs.
E.g. we consider a program $\Sigma$ to be a sequence of function
definitions of the form $F\,\Gamma = t \tin \sigma$, we have to
parameterise every judgement by $\Sigma$ and require that $\Gamma
\vdash_\Sigma t \tin \sigma$ for the definition to be a wellformed
extension of $\Sigma$. We also have to introduce a rule for
function-application which can just be translated into an iterated
let-expression.

\section{Conclusions and further work}
\label{sec:concl}

We have introduced a language for finite quantum programs which
uniformly extends a finitary classical language. The classical part of
our language may be of interest for its own sake, as it introduces
a natural way to compile functional terms into space efficient
reversible circuits, due to no unnecessary garbage. This
uniformity is one of the main design principles of our language,
which, we hope, makes it a natural vehicle to express quantum
programming and to develop \emph{quantum thinking}.

We are currently implementing a compiler for QML in Haskell. The
compiler produces a representation of quantum circuits which can be
simulated (inefficiently, of course) by our own simulator or by
using a standard simulator for quantum gates.

There are other design ideas for quantum programming languages. A
potential criticism of our approach is that we leave contractions
implicit, which is an operation which depends on the choice of
basis. However, our type assignment system clearly fixes the places
where contractions have to happen, and moreover, and we believe more
importantly, it fixes the places where projections, or
\emph{tracing}, is happening. A central feature of any
quantum programming language seems to be \emph{control of decoherence}.

Having noted this, it seems that decoherence is something you
always want to minimise. It is straightforward to design an
inference algorithm which infers weakenings
$t^{\dom{\G}}$ such that decoherence is minimised. Maybe this should
be the default, which can be overridden if the programmer wants to
enforce measurement.

We would like to have an orthogonality judgement which is complete with
respect to the denotational semantics. One of the referees commented
that we would need an inner product judgement to achieve this. We plan
to explore this proposal in future work.

The restriction that \ensuremath{\mathbf{case}^\circ} is only allowed for balanced coproducts
is a direct reaction to the comments of the same referee who
pointed out that our previous approach, which involved padding the
data, is problematic.  Indeed, this problem seems unfixable; if we
branch over $\QQ{1}\otimes\Qbit$ the garbage which is created by
padding may indirectly measure the qubit we are branching over.
Consequently, this approach would not be compositional, and hence
should be rejected.  The inability to deal with quantum control over
arbitrary coproducts is a consequence of the fact that while we deal
with quantum data and control, the structure, i.e. the memory
allocation, of our data is classical. One way to overcome this
limitation would be to use an operational semantics which employs a
quantum memory allocation. Such a semantics would have to exploit an
infinite state space, and it is questionable whether such a system is
physically plausible. Another direction, which seems more feasible,
would be to index quantum structures by classical values at compile
time. 

We have some doubts as to whether the understanding of general
recursion and partiality in quantum programming is essential, because
partiality is only interesting for systems with infinite state spaces.
Moreover, it is not clear how to observe the termination of such a
hypthothetical quantum system of unknown runtime without disturbing
the computation.

Higher order programming would be a worthwhile addition to reflect the
way many quantum algorithms are presented: e.g. the Quantum Fourier
Transform can be parameterised by a function on quantum words.
Recently, Selinger investigated this problem \cite{selinger:qsem}
and it seems that
currently no canonical higher order structure on $\Super$ is known.
We are investigating whether the category of presheaves over $\Super$
would provide a sound denotational model for higher order quantum
computation. This semantics would employ Day's construction to 
interpret tensor products. 

Another line of work is to reap the benefits of the fact that our
language uses high level constructs, and develop high level
reasoning principles for QML programs. To achieve this, our next goal
is to give a direct translation of QML to superoperators which
factors through the $\FQC$ semantics presented here. This
translation will be based on the implementation of superoperators
using arrows \cite{hughes:arrows} in Haskell \cite{alti:qeff-draft}.
A direct consequence of this construction is that the translation
presented here is compositional with respect to the extensional
equality.

In joint work with Sabry and Vizzotto we are currently developing an
equational theory for QML, an algebra for quantum programming, which
is sound and complete, with respect to the denotational semantics suggested 
here. Since the completeness proof relies on \emph{inverting evaluation},
such a proof also gives rise to normalisation; exploiting
the approach developed in \cite{alti:flops04} for a classical system.

\section*{Acknowledgements}
We would like to acknowledge interesting discussions on
the subject of this paper with  Slava Belavkin, Martin Hofmann, Conor
McBride, Alex Simpson and Thomas Streicher. Amr Sabry and Juliana Vizotti
provided extensive feedback on previous drafts of this paper. Peter Selinger
pointed out a serious flaw in the definition of \ensuremath{\mathbf{case}^\circ} and refuted our
conjecture that strict maps classify monos in $\Super$. We would like
to thank the anonymous referees for their valuable feedback, especially one
of the referees, who provided very detailed and extremely useful
technical comments on our work.

\bibliographystyle{plain}
\bibliography{local}

\end{document}